\documentclass[letterpaper,twocolumn,english,superscriptaddress]{revtex4}

\usepackage{lineno}
\usepackage[T1]{fontenc}
\usepackage[latin9]{inputenc}
\usepackage{amsmath}
\usepackage{amssymb}
\usepackage{graphicx}
\usepackage{dcolumn}
\usepackage{verbatim}
\usepackage{slashed}

\makeatletter

\pdfpageheight\paperheight
\pdfpagewidth\paperwidth

\usepackage{babel}

\makeatother

\usepackage{babel}

\usepackage{tikz,xcolor,hyperref}
\definecolor{lime}{HTML}{A6CE39}
\DeclareRobustCommand{\orcidicon}{
	\begin{tikzpicture}
	\draw[lime, fill=lime] (0,0) 
	circle [radius=0.16] 
	node[white] {{\fontfamily{qag}\selectfont \tiny ID}};
	\draw[white, fill=white] (-0.0625,0.095) 
	circle [radius=0.007];
	\end{tikzpicture}
	\hspace{-2mm}
}

\foreach \x in {A, ..., Z}{%
	\expandafter\xdef\csname orcid\x\endcsname{\noexpand\href{https://orcid.org/\csname orcidauthor\x\endcsname}{\noexpand\orcidicon}}
}

\begin{document}
\title{Measurements of the lightest hypernucleus ($\mathrm{^3_{\Lambda}H}$): progress and perspective}

\author{Jinhui Chen}
\affiliation{Key Laboratory of Nuclear Physics and Ion-beam Application (MoE), Institute of Modern Physics, Fudan University, Shanghai 200433, China}
\affiliation{Shanghai Research Center for Theoretical Nuclear Physics, NSFC and Fudan University, Shanghai 200438, China}

\author{Xin Dong}
\affiliation{Lawrence Berkeley National Laboratory, Berkeley, California 94720, USA}

\author{Yu-Gang Ma}
\affiliation{Key Laboratory of Nuclear Physics and Ion-beam Application (MoE), Institute of Modern Physics, Fudan University, Shanghai 200433, China}
\affiliation{Shanghai Research Center for Theoretical Nuclear Physics, NSFC and Fudan University, Shanghai 200438, China}

\author{Zhangbu Xu}
\affiliation{Brookhaven National Laboratory, Upton, New York 11973, USA}

\maketitle

\paragraph{Abstract}
The hyperon-nucleon ($Y$-$N$) interaction is important for the description of the equation-of-state of high baryon density matter. Hypernuclei, the cluster object of nucleons and hyperons, serve as cornerstones of a full understanding of the $Y$-$N$ interaction. Recent measurements of the lightest known hypernucleus, the hypertriton's ($\mathrm{^3_{\Lambda}H}$) and anti-hypertriton's ($\mathrm{^3_{\bar{\Lambda}}\bar{H}}$) lifetime, mass and $\Lambda$ separation energy have attracted interests on the subject. Its cross section and collective flow parameters have also been measured in heavy-ion collisions, which have revealed new features on its production mechanism. In this article we summarise recent measurements of $\mathrm{^3_{\Lambda}H}$, focusing on the heavy-ion collisions. We will discuss their implications for the $\mathrm{^3_{\Lambda}H}$ properties and the constrains on the $Y$-$N$ interaction models.

\section{Introduction}
Although the structure of nuclei in nature is determined by the strong force that bind neutrons and protons into the clusters, and we have a good understanding of the quantum chromodynamics (QCD) theory at the fundamental level~\cite{Gross:1973id,Politzer:1973fx}, the theory itself is not always directly applicable to nuclear structure. It remains one of the scientific quests on whether the fundamental interactions that are basic to the structure of matter could be fully understood~\cite{NSACLRP:2015qub}. As the energy scale increases in nuclear reactions and/or the energy density increases in stars, strangeness quantum number may come into play~\cite{Itoh:1970uw}. Hypernuclei are bound state of normal nuclei with an additional strange baryon, such as $\Lambda$, $\Sigma$, $\Xi$ and $\Omega$ hyperons. Such nuclear objects have been the natural tool to study the hyperon-nucleon ($Y$-$N$) interaction since the discovery of the first element~\cite{danysz1953delayed} exactly seven decades ago. The $Y$-$N$ interaction is theoretical predicted to be the important ingredient to describe the equation-of-state of astrophysical objects such as neutron stars~\cite{Lattimer:2004pg,Tolos:2020aln}, which, depending on the strength of the interaction, might be a hyperon star, might be composed of strange quark matter, or might have a kaon condensate at its core~\cite{Lattimer:2004pg}. Because hyperons are not stable in the Vacuum, it is challenging to operate direct hyperon-nucleon scattering experiments in the way of electron scattering to probe the nuclei structure. To date, hypernucleus still represents a practical laboratory ideal for studying the $Y$-$N$ interaction~\cite{Chrien:1989xeq,Hashimoto:2006aw}. 

The lowest mass bound hypernucleus is the hypertriton ($\mathrm{^3_{\Lambda}H}$), which consists of a $\rm \Lambda$, a proton and a neutron. It is the most copiously produced hypernucleus in heavy-ion reaction, where is the main scope of current paper. By colliding heavy nuclei at relativistic energies scientists are able to test the nature of nuclear matter at high temperature and density, to produce conditions similar to those thought prevalent in the early universe, and to search for previously unstudied states of nuclear matter~\cite{Shuryak:1980tp}. Relativistic heavy-ion collisions create a hot and dense phase of matter containing approximately equal number of quarks and antiquarks, where they are free to move throughout the volume of the nuclear collision region. This phase of matter persists for only a few times $10^{-23}$ seconds, then cools and transitions into a less excited phase comprised of mesons, and baryons, including the occasional (anti)nucleus or (anti)hypernucleus. Thus these collisions could be used to explore fundamental physics involving nuclei, hypernuclei~\cite{Chen:2018tnh}. With the added strangeness degree of freedom $\mathrm{^3_{\Lambda}H}$ is expected to possess different properties from normal nucleus with similar atomic mass number, such as triton or helium. Therefore, measurements of  $\mathrm{^3_{\Lambda}H}$ binding energy, excitation energies for particle-bound state, spins, lifetimes, and decay branching ratios will allow us to infer the $Y$-$N$ interaction from the simplest hypernuclear cluster~\cite{Gal:2016boi}.

\begin{figure}
\includegraphics[width=8.cm]{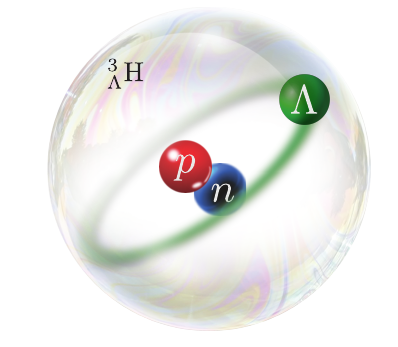}
\caption{\label{fig:h3l-structure}(Color online) Sketch of the $\mathrm{^3_{\Lambda}H}$ structure. Although proton, neutron and $\Lambda$ by themselves are color neutral, they are painted here by different color for viewing.}
\end{figure}

\section{The lifetime and $\Lambda$ separation energy of $\mathrm{^3_{\Lambda}H}$: status and progress}
\subsection{Status in the early days}

The $\Lambda$ separation energy of $\mathrm{^3_{\Lambda}H}$, the $B_{\Lambda}$ representing the energy needed to remove $\Lambda$ from the deuteron core of $\mathrm{^3_{\Lambda}H}$, is defined as
\begin{equation}
  B_{\Lambda}\equiv M_d + M_{\Lambda} - M_{\mathrm{^{3}_{\Lambda}H}},
\end{equation}
which is rather small compared to the normal nuclei or heavy hypernuclei~\cite{Juric:1973zq,STAR:2019wjm,ALICE:2022rib}.
Thus it was argued that the $\Lambda$ is loosely bound outside the deuteron core in the $\mathrm{^3_{\Lambda}H}$~\cite{1992.JPG.18.339}, as shown in the Fig.~\ref{fig:h3l-structure}. For example, taking the frequently cited value of $B_{\Lambda}$ = 0.13 $\pm$ 0.05 MeV~\cite{Juric:1973zq} will result in a large separation of the $\Lambda$ from the deuteron of about 10 fm~\cite{Hildenbrand:2019sgp}. Consequence of such structure is that the lifetime of $\mathrm{^3_{\Lambda}H}$, the $\tau(\mathrm{^3_{\Lambda}H})$ should be very close to the free $\Lambda$ lifetime ($\tau_{\Lambda}$), whose value is $\tau_{\Lambda}=263.2 \pm2.0~ps$~\cite{ParticleDataGroup:2022pth}. However, measurements of $\tau(\mathrm{^3_{\Lambda}H})$ using nuclear emulsions and helium bubble chambers showed a large spread of $\tau(\mathrm{^3_{\Lambda}H})$ value, ranging from $\approx$ 0.35$\tau_{\Lambda}$ to 1.08$\tau_{\Lambda}$ with very large statistical uncertainty~\cite{1964.PR.136.B1803,1968.PRL.20.819,1969.PR.180.1307,1970.NPB.16.46,1970.PRD.1.66,1973.NPB.67.269}. This raised a puzzle in the field and called for more precise measurements of the $\mathrm{^3_{\Lambda}H}$ lifetime to clarify the situation~\cite{Davis:2005mb,Dalitz:2005mc}. 

\subsection{Lifetime measurements in HIC}

\begin{figure}
\includegraphics[width=8.0cm]{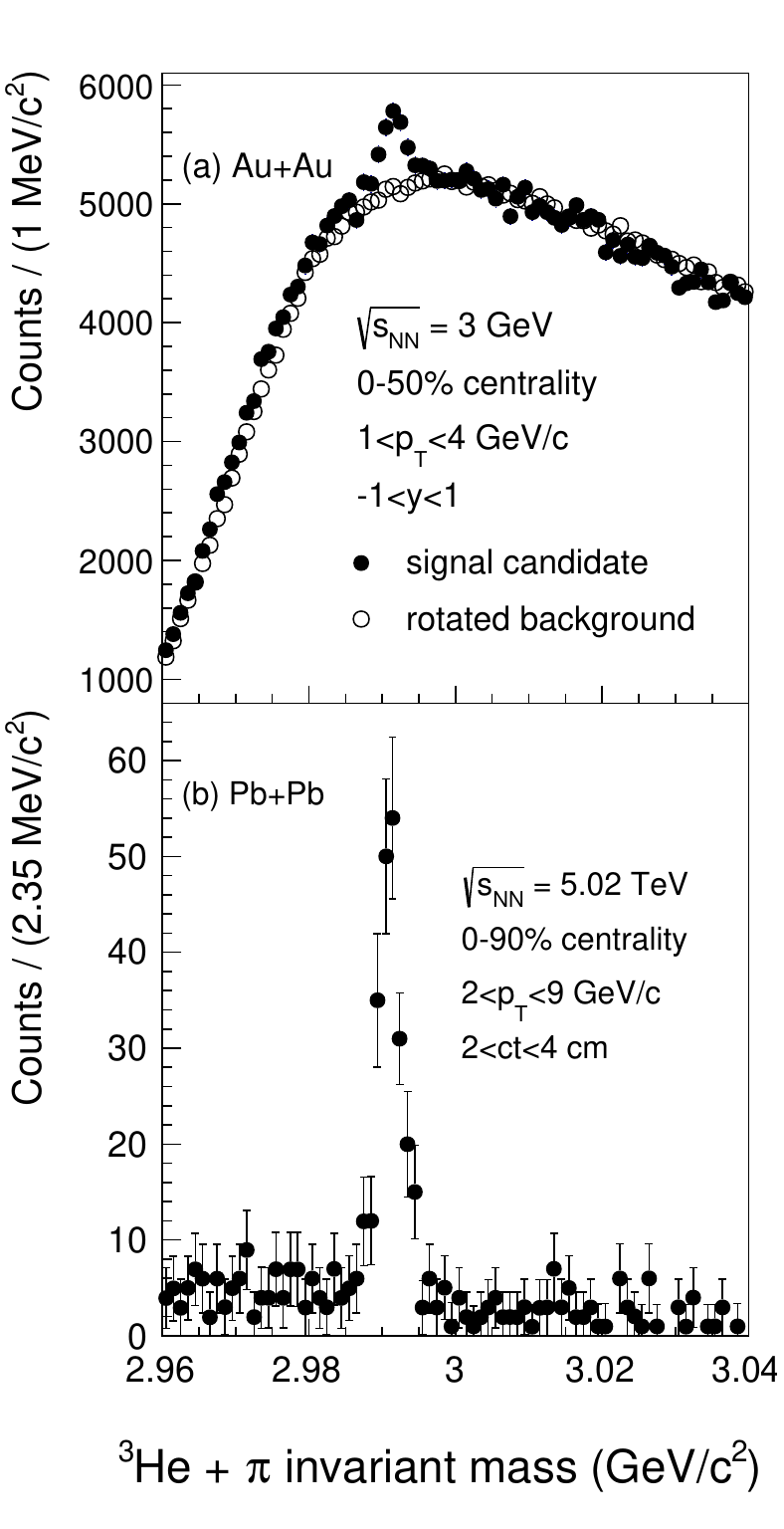} 
\caption{\label{fig:inv-mass-ctau}(Color online) Invariant mass distributions of $\mathrm{^3_{\Lambda}H}$ candidate reconstructed by $\rm ^{3}He + \pi^{-}$ with panel (a) from Au+Au collisions at $\rm \sqrt{s_{NN}}=$ 3 GeV~\cite{STAR:2021orx} and panel (b) from Pb+Pb collisions at $\rm \sqrt{s_{NN}}=$ 5.02 TeV~\cite{ALICE:2022rib}. Black circles in both panels represent the signal distributions, while open circles in panel (a) are combinatorial background distribution built with rotational daughter track algorithm.}
 \end{figure}

The technology to measure the $\tau(\mathrm{^3_{\Lambda}H})$ in heavy-ion collisions is different from those applied in early days. In helium bubble chamber or nuclear emulsion experiments, $\mathrm{^3_{\Lambda}H}$ was observed by analyzing the events produced in the interaction of $K^-$ both in flight and at rest with the nuclei of the sensitive layers of visualizing detectors. In heavy-ion collisions, they are measured by detecting their mesonic decay via the topological identification of secondary vertices and the analysis of the invariant mass distributions of decay particles. Taking the $\mathrm{^3_{\Lambda}H}$ decays to $\rm ^3He$ and $\pi$ in Relativistic Heavy Ion Collider, the STAR experiment as an example, the main STAR detectors used to identify the candidate $\rm ^3He$ and $\pi$ are the Time Projection Chamber (TPC) combined with the Time of Flight detector (TOF), by the mean energy loss per unit track length ($\mathrm{\langle dE/dx \rangle}$) in the TPC gas and the speed ($\beta$) determined from TOF measurements. The $\mathrm{^3_{\Lambda}H}$ is then measured via the invariant mass of $\rm ^3He$ and $\pi$. With reconstruction of the $\mathrm{^3_{\Lambda}H}$ weak decay vertex up to a few cm away from the collision vertex, together with the high precision tracking and particle identification capabilities of the STAR experiment, the invariant mass of each $\mathrm{^3_{\Lambda}H}$ candidate is calculated with a good signal to background ratio. Fig.~\ref{fig:inv-mass-ctau} (a) shows such a distribution in Au+Au collisions at $\rm \sqrt{s_{NN}}=$ 3 GeV~\cite{STAR:2021orx}. To improve the signal to background ratio, machine learning techniques have also been applied in the new data analysis. Taking the result of Ref.~\cite{ALICE:2022rib} in Large Hadron Collider, the ALICE experiment as an example where the detector apparatus is similar to the STAR experiment: cut criteria to optimize the signal are performed with a gradient-boosted decision tree (BDT) classifier implemented by the XG-BOOST library and trained on a dedicated Monte Carlo simulated event sample~\cite{ALICE:2022rib}. After such trainings, the selection is based on the BDT score, defining a threshold that maximizes the expected signal to background ratio~\cite{ALICE:2022rib}. Fig.~\ref{fig:inv-mass-ctau} (b) shows such a distribution in Pb+Pb collisions at $\rm \sqrt{s_{NN}}=$ 5.02 TeV with almost background free mainly due to the excellent performance of the ALICE Inner Tracking System and the TPC detector~\cite{ALICE:2022rib}.

\begin{figure}[htbp]
\includegraphics[width=8.0cm]{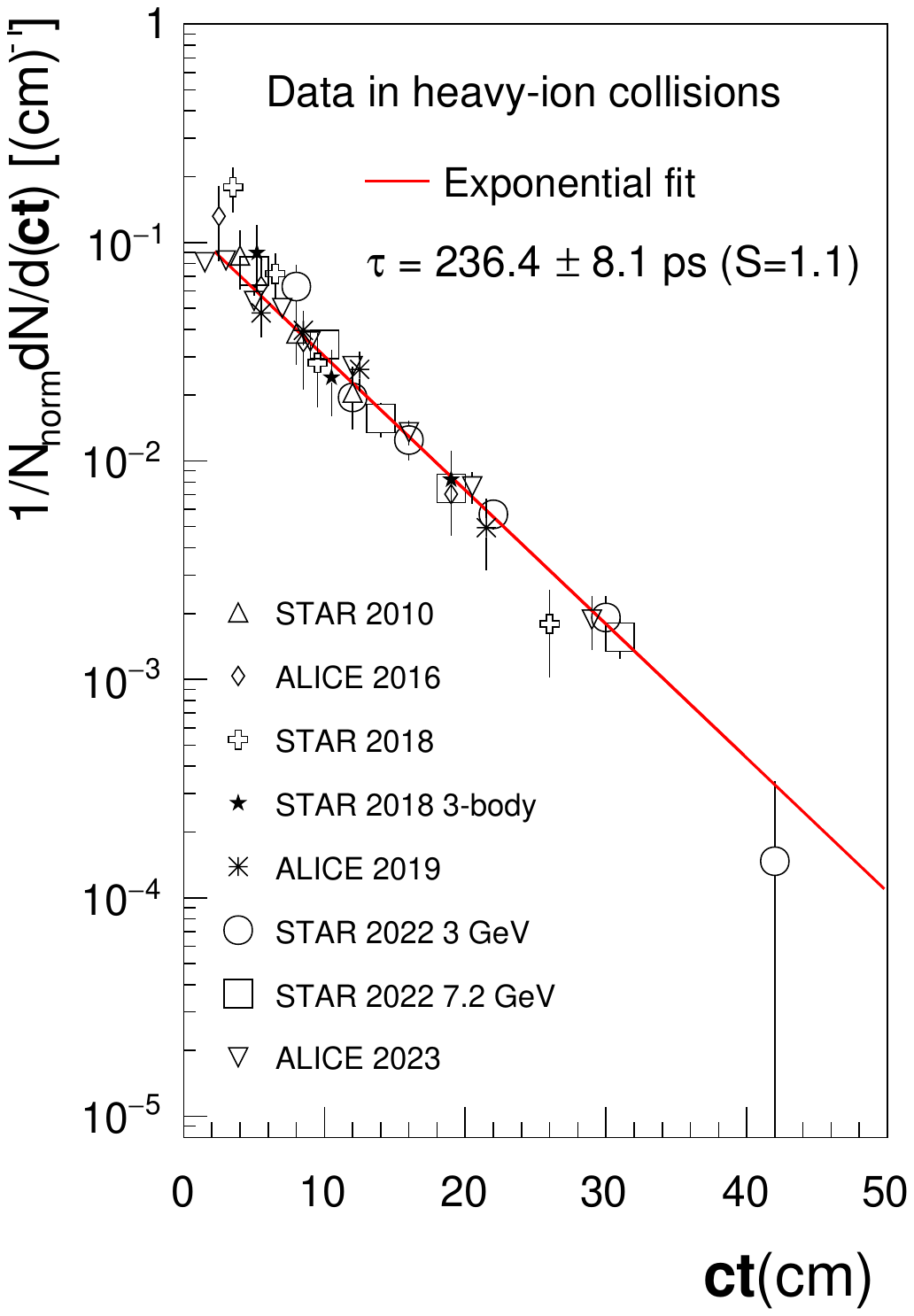}
\caption{\label{hic-ctau-fit}(Color online) The normalized yield as a function of the proper decay length $ct$ for $\mathrm{^3_{\Lambda}H}$ in heavy-ion collisions. The $\tau$ parameter of the exponential fit is determined by a common fit to the 8 individual measurements with the same slope parameter, and the associated uncertainty is scaled by S=1.1.}
\end{figure}

To determine the lifetime, the $\mathrm{^3_{\Lambda}H}$ yields are analyzed in differential proper decay length intervals according to the equation 
\begin{equation}
N(t) = N(0) \mathrm{exp}(-t/\tau),
\end{equation} 
where $t=l/(\beta\gamma c)$, $\beta\gamma c = p/m$, $l$ is the measured decay distance, $p$ is the particle momentum, $m$ is the particle mass, and $c$ is the speed of light. The raw count in different $ct$ bin was extracted from the invariant mass distribution [c.f. Fig.~\ref{fig:inv-mass-ctau}]. Then the distribution was corrected for the detector acceptance, tracking efficiency and selection efficiency as a function of $ct$. Fig.~\ref{hic-ctau-fit} shows such distributions in heavy-ion collisions. The lifetime is then extracted by a $\chi^2$ fit of each measurement with the exponential function.

\begin{figure}
\includegraphics[width=8.8cm]{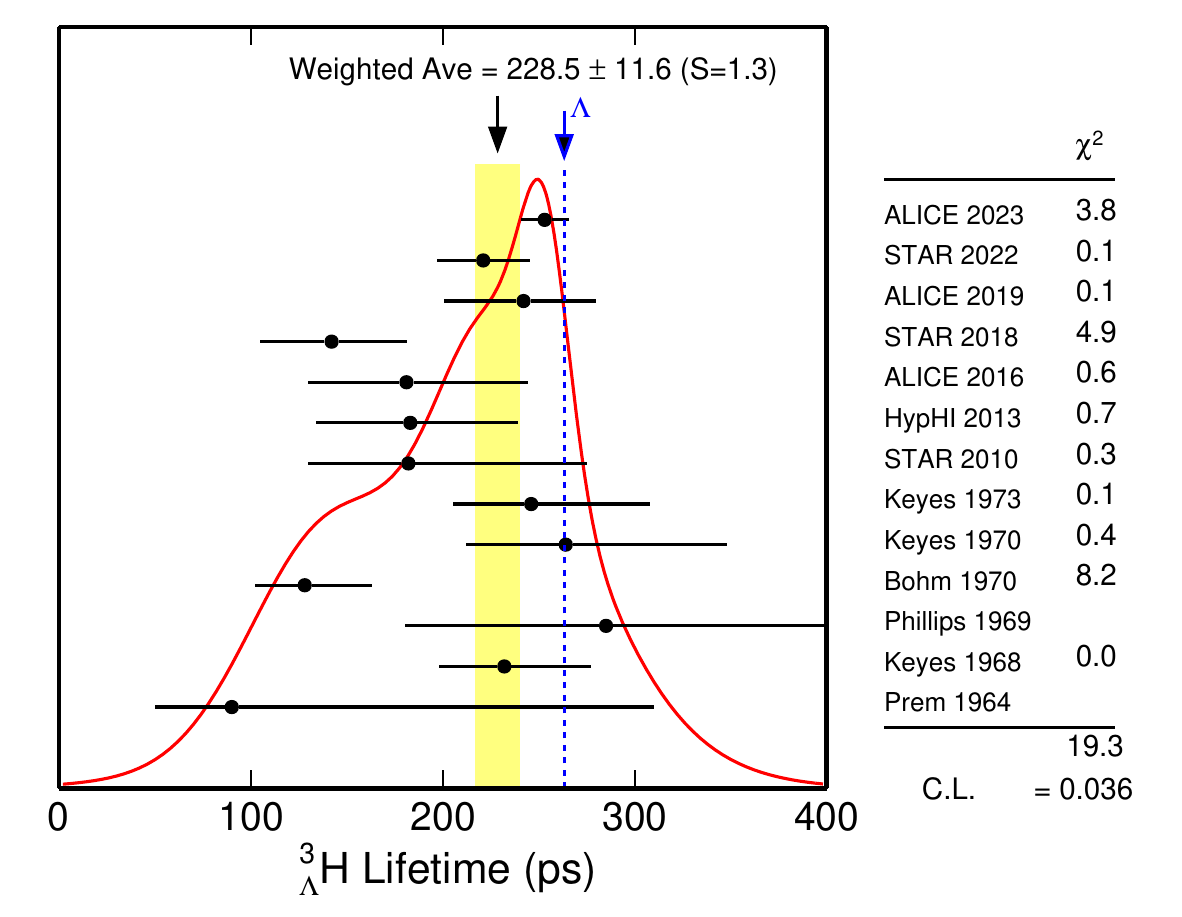}
\caption{\label{fig:lifetime-data}(Color online) A typical ideogram of the $\tau(\mathrm{^3_{\Lambda}H})$. The arrow at the top shows the position of the weighted average, while the width of the shaded pattern shows the error in the average after scaling by the factor $S$. The column on the right gives the $\chi^2$ contribution of each of the experiments. All data are cited in reference~\cite{1964.PR.136.B1803,1968.PRL.20.819,1969.PR.180.1307,1970.NPB.16.46,1970.PRD.1.66,1973.NPB.67.269,STAR:2010gyg,Rappold:2013fic,ALICE:2015oer,STAR:2017gxa,ALICE:2019vlx,STAR:2021orx,ALICE:2022rib}. The dashed line is the $\tau_{\Lambda}$ for reference~\cite{ParticleDataGroup:2022pth}.}
\end{figure}

Figure~\ref{fig:lifetime-data} shows a collection of the world data on $\tau(\mathrm{^3_{\Lambda}H})$ measurements. In comparison with the lifetime of free $\Lambda$, the $\mathrm{^3_{\Lambda}H}$ data is ranging from $\sim$50\% to $\sim$100\% of free $\Lambda$ with considerable uncertainties. An average value of $228.5~ps$ is obtained by taking the average with 1/($\sigma$)$^2$ as the weight for each measurement, where $\sigma$ represents the total experimental uncertainty (statistical and systematical uncertainties added in quadrature if the latter is available). The associated uncertainty on the average value is $11.6~ps$, which has been scaled by the $S\equiv\sqrt{\chi^2/(N-1)}$ value of 1.3. Fig.~\ref{fig:lifetime-data} also shows the PDG ideogram distribution, a probability distribution weighted by $1/\sigma$ while $\sigma$ is the individual experimental data uncertainty. Investigation on the difference between the ideogram curve and measurements may offer information for potentially additional systematics which are not covered by the measurement.

Since measurements in heavy-ion collisions are carried out as a function of proper decay length, it allows for a common fit of the distributions to obtain an average value of $\tau(\mathrm{^3_{\Lambda}H})$ in heavy-ion collisions. 
In this analysis, we fit the individual distributions of heavy-ion data simultaneously with a same $\tau$ parameter and free normalization parameter for each distribution. The data are described well by the aforementioned exponential function with a probability of 0.2, shown as the red line in Fig.~\ref{hic-ctau-fit}. The $\chi^2/\it NDF$ = 1.22 is slightly larger than unit, which is mainly driven by the data points at lowest $ct$. We observe that the STAR data point at lowest $ct$ (the open cross) appears to be systematically higher than the exponential function with 2.47$\sigma$ while the ALICE data point at lowest $ct$ (the open triangle) is lower than the exponential function with 2.10$\sigma$. The $\tau(\mathrm{^3_{\Lambda}H})$ = $236.4 \pm 8.1~ps$ is the value from a common fit of data in heavy-ion collisions, where the uncertainty has been scaled by the $S$ value of 1.1. The lifetime of $\mathrm{^3_{\Lambda}H}$ from a common fit of heavy-ion collisions data solely is consistent with the average value of world data including early nuclear emulsion measurements and bubble chamber measurements. Our number is $\sim$10\% smaller than the value of $\tau_{\Lambda}$~\cite{ParticleDataGroup:2022pth}.

\subsection{$\Lambda$ separation energy}

\begin{figure}[htbp]
\includegraphics[width=0.48\textwidth]{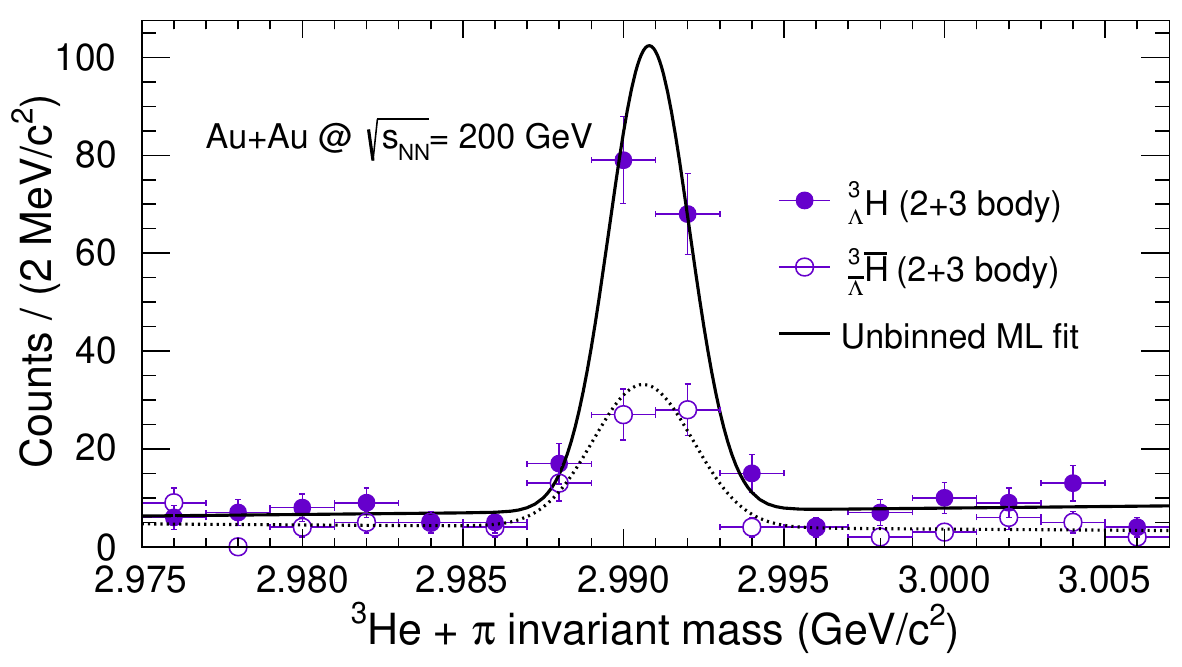}
\caption{\label{fig:bindingE_mass}(Color online) The invariant mass distributions of $\mathrm{^3_{\Lambda}H}$ reconstructed via the charged two-body and three-body decay channels $\mathrm{^3_{\Lambda}H \rightarrow ^3He + \pi^-}$, $\mathrm{^3_{\Lambda}H \rightarrow d + p + \pi^-}$ and their corresponding charge-conjugated particles $\mathrm{^3_{\bar{\Lambda}}\bar{H}}$ in Au+Au collisions at $\mathrm{\sqrt{s_{NN}}}$ = 200 GeV by the STAR experiment~\cite{STAR:2019wjm}. Points are data and curves represent a fit using a Gaussian function plus a linear polynomial for illustration purpose. Mass extraction procedure should be referred to the text of Ref.~\cite{STAR:2019wjm}.}
\end{figure}

The measurement of $B_{\Lambda}$ requires a precise determination of the rest mass of $^{3}_{\Lambda}$H. In heavy-ion collisions, $^{3}_{\Lambda}$H particle is often reconstructed via its two- or three-body decays involving charged particles by calculating the invariant mass of $^{3}_{\Lambda}$H through its decay daughters. The determination of $M({\mathrm{^{3}_{\Lambda}H}})$ will then require precision momentum calibration of the decay charged daughter tracks. The $\Lambda$ with a mass of 1115.683 $\pm$ 0.006 MeV~\cite{ParticleDataGroup:2022pth} whose decay has a similar decay topology as the $\mathrm{^3_{\Lambda}H}$ decay, therefore is often used to calibrate the experimental momentum distortion correction. Fig~\ref{fig:bindingE_mass} shows one of the mass distributions in Au+Au collisions at $\mathrm{\sqrt{s_{NN}}}$ = 200 GeV by the STAR experiment~\cite{STAR:2019wjm}. 
Due to the installed of high-precision tracking detector which was not applied in the Fig.~\ref{fig:inv-mass-ctau} (a) analysis, the invariant mass distributions are reconstructed with a low level of background, which mainly originates from combinatorial contamination and particle misidentification. In the STAR measurement, the invariant mass distributions were then fitted by a Gaussian function plus a linear polynomial using the unbinned maximum likelihood method. The mass parameters are extracted through the peak positions of the invariant mass distributions. An average of mass values from two-body and three-body decay channels weighted by the reciprocal of squared statistical uncertainties is the final result. One can perform the same average on $\mathrm{^3_{\Lambda}H}$ and its antiparticle to improve the precision~\cite{STAR:2019wjm}.

\begin{figure}[htbp]
\includegraphics[width=0.5\textwidth]{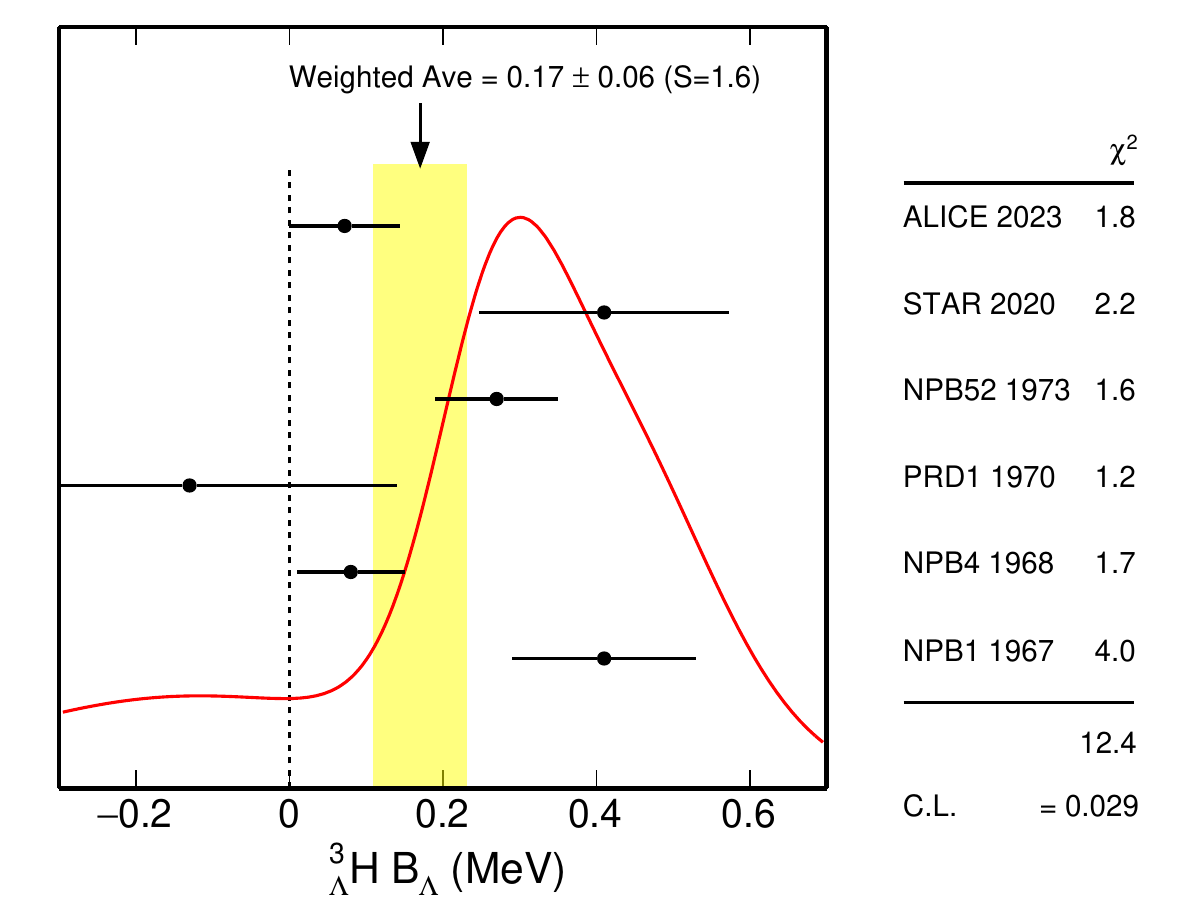}
\caption{\label{fig:bl-data}(Color online) 
Similar to Fig.~\ref{fig:lifetime-data} but for measurements of the $\Lambda$ separation energy of $\mathrm{^3_{\Lambda}H}$~\cite{Gajewski:1967ruj,Bohm:1968qkc,1970.PRD.1.66,Juric:1973zq,STAR:2019wjm,ALICE:2022rib}. The arrow at the top shows the position of the weighted average, while the width of the shaded pattern shows the error in the average after scaling by the factor $S$. The column on the right gives the $\chi^2$ contribution of each of the experiment. The dashed line represents $\rm B_{\Lambda}=$0 for reference.}
\end{figure}

Figure~\ref{fig:bl-data} summarizes the experimental measurements of $B_{\Lambda}$ of $\mathrm{^3_{\Lambda}H}$ from early nuclear emulsion experiments and recent heavy-ion collisions~\cite{Gajewski:1967ruj,Bohm:1968qkc,1970.PRD.1.66,Juric:1973zq,STAR:2019wjm,ALICE:2022rib}. The original data from early experiments have been corrected for the updated $M_d$ and $M_{\Lambda}$ since more than half a century ago~\cite{Liu:2019mlm}. The modern heavy-ion experimental data contain both statistical and systematical uncertainties which are added in quadrature in the presented error bars. The yellow band depicts a global average of all data points which yields $B_{\Lambda} = 0.17 \pm 0.06$\,MeV, where the uncertainty has been scaled by the $S$ = 1.6. Please note that the $S$ value of 1.6 from the average is rather large, indicating a relative large spread of experimental data points. 

The red line in Fig.~\ref{fig:bl-data} shows the ideogram presentation of the data points which also shows a large spread of the values. Despite the small systematical error assigned to the nuclear emulsion measurements~\cite{Davis:2005mb}, it was shown that for $p$-shell hypernuclei a discrepancy exists in the range of 400 to 800 keV between nuclear emulsion data~\cite{Davis:2005mb} and those obtained with ($\pi^+,K^+$) reaction~\cite{Gogami:2015tvu}. Recalibration of the spectrometer data with respect to different measurements, for example, the FINUDA data~\cite{Botta:2016kqd} reduced the difference, there remained substantial differences between early nuclear emulsion studies and recent spectrometer data with a significant spread of a few hundreds keV for individual hypernuclei. Similar situation is present in the heavy-ion collisions data. A full understanding on the systematics is called for.  

\begin{figure}[htbp]
\includegraphics[width=0.5\textwidth]{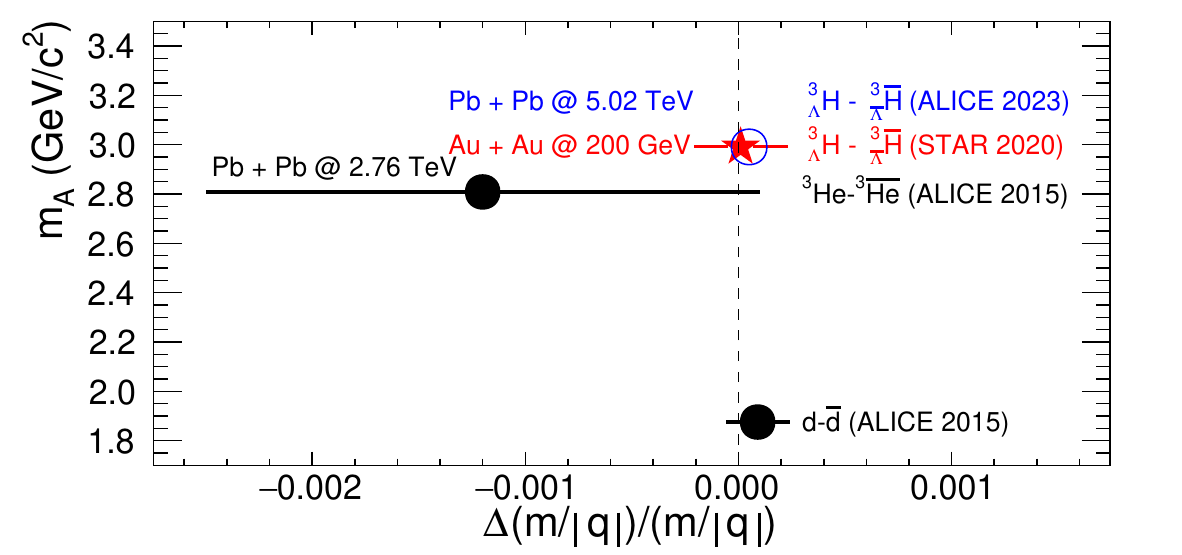}
\caption{\label{fig:massdiff} Data~\cite{ALICE:2015rey,STAR:2019wjm,ALICE:2022rib} of the mass-over-charge ratio differences for light nuclei and hypernuclei from heavy-ion collisions compared with CPT invariance expectation (dotted line). Error bars represent the sum  in quadrature of the statistical and systematical uncertainties.}
\end{figure}
Benefit from the excellent resolution of the anti-nuclei mass measurements in high-energy heavy-ion collisions~\cite{STAR:2011eej,ALICE:2015rey,ALICE:2017jmf,STAR:2019wjm,ALICE:2022rib}, one can measure the mass difference of $\mathrm{^3_{\Lambda}H}$ and $\mathrm{^3_{\bar{\Lambda}}\bar{H}}$ to probe the matter-antimatter symmetry pertaining to the binding of strange and anti-strange quarks in nuclei. The fundamental symmetry of nature, the CPT theorem which exchanges particles with anti-particles, implies that all physics laws are the same under the simultaneous reversal of charge, reflection of spatial coordinates and time inversion. Although measurements of mass difference between baryon-antibaryon had reached a very high precision with protons and anti-protons~\cite{Gabrielse:1999kc}, extension the measurement from (anti-)proton to (anti-)nuclei allows one to probe any difference in the interactions between nucleons and anti-nucleons encoded in the (anti-)nuclei masses. This force is a remnant of the underlying strong interaction among quarks and gluons which cannot yet be directly derived from QCD. Fig.~\ref{fig:massdiff} shows the mass-over-charge ratio difference, $\frac{\Delta{m/|q|}}{m/|q|} = \frac{m_{\rm A} - m_{\rm \bar{A}}}{m_{\rm A}}$, for $d$, ${\rm ^3He}$ in Pb+Pb collisions at $\rm \sqrt{s_{NN}}=$ 2.76 TeV~\cite{ALICE:2015rey} and $\mathrm{^3_{\Lambda}H}$ in Au+Au collisions at $\rm \sqrt{s_{NN}}=$ 200 GeV and Pb+Pb collisions at $\rm \sqrt{s_{NN}}=$ 5.02 TeV~\cite{STAR:2019wjm,ALICE:2022rib}. Systematical uncertainties can be reduced by performing mass differences, rather than absolute masses measurements. Within current uncertainties, all data presented in Fig.~\ref{fig:massdiff} are consistent with zero, the dashed line in Fig.~\ref{fig:massdiff} from CPT invariance expectation. Data from heavy-ion collisions confirm CPT invariance in the sector of light nuclei and light hypernuclei.

\section{The production yields and collective flow of $\mathrm{^3_{\Lambda}H}$ in HIC}

\begin{figure}[htbp]
\includegraphics[width=0.46\textwidth]{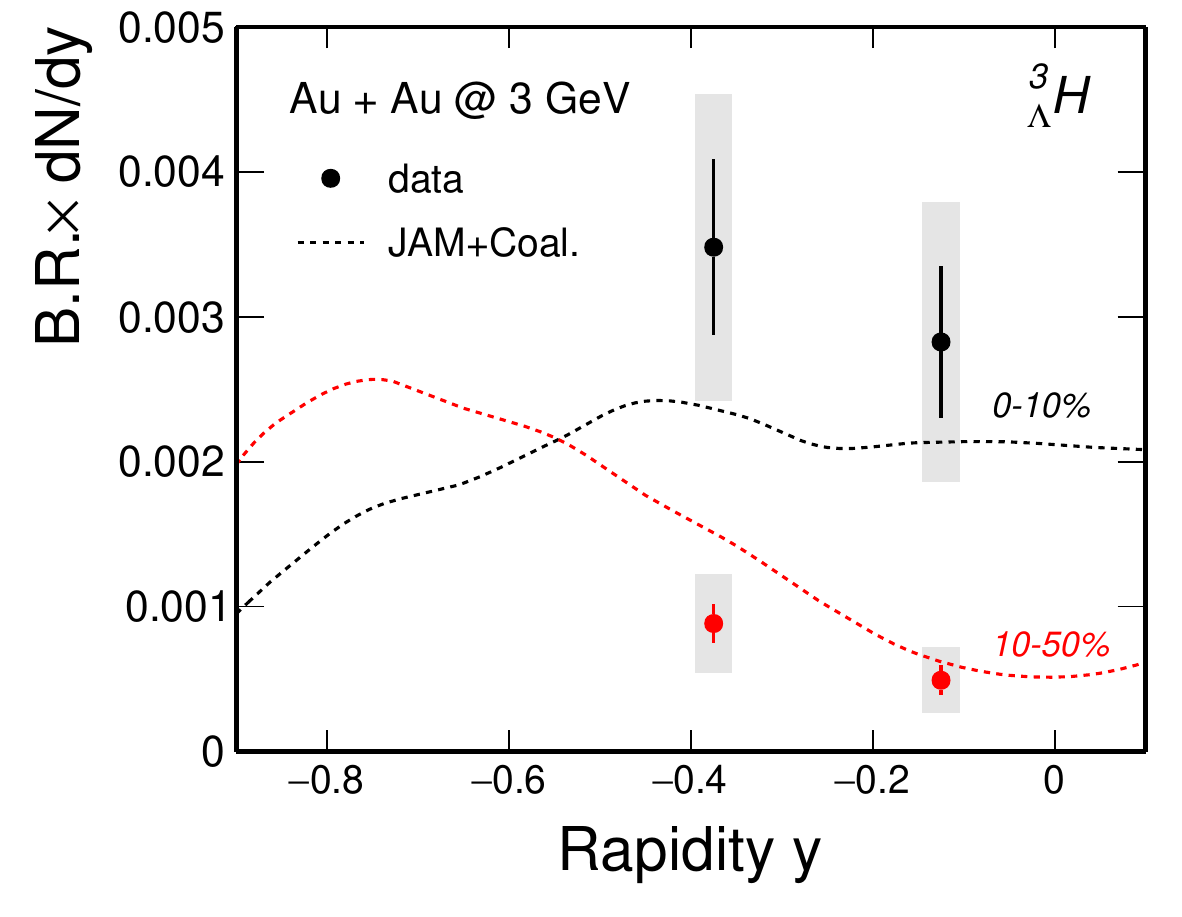}
\includegraphics[width=0.46\textwidth]{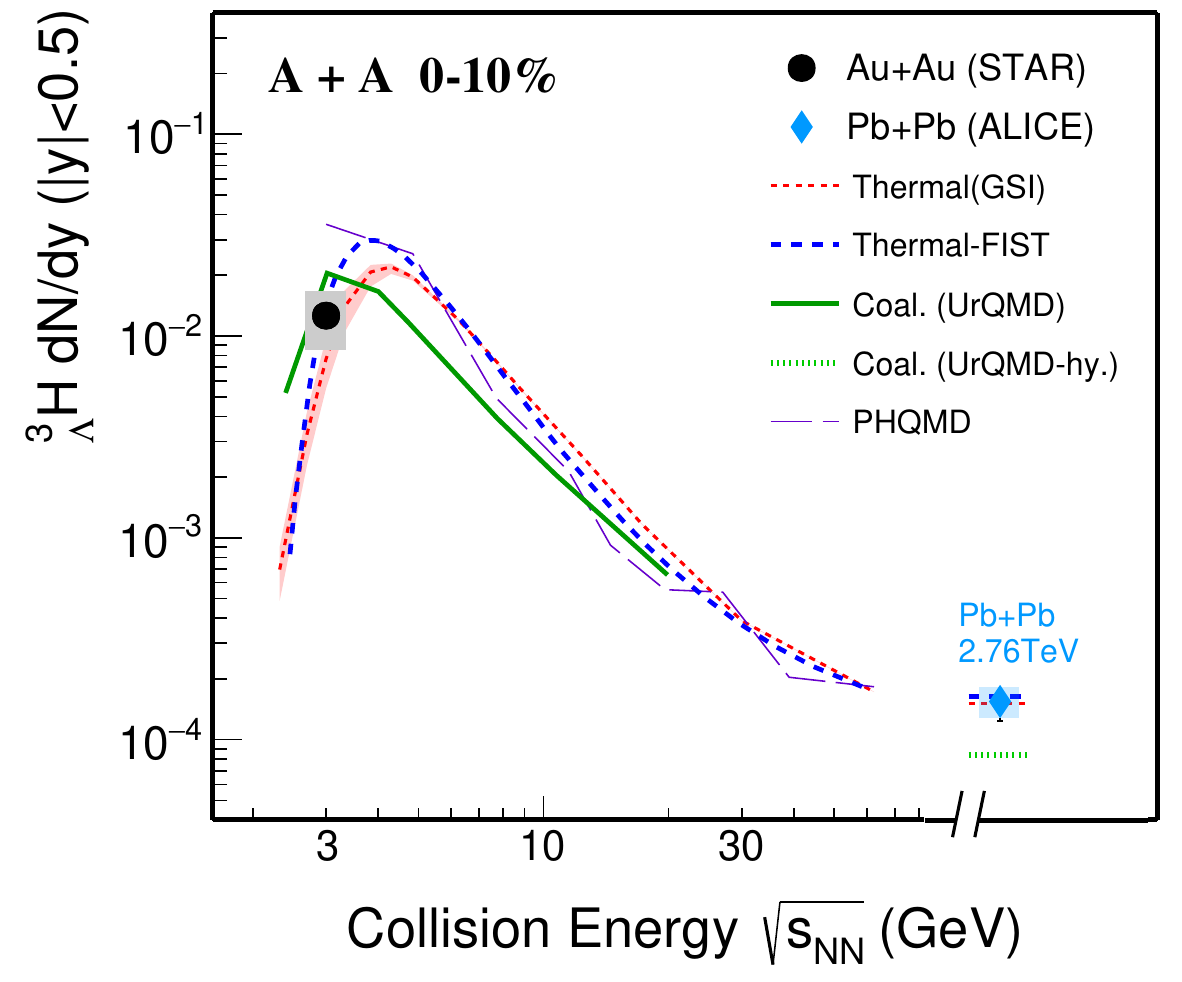}
\caption{\label{fig:yield-data}(Color online) Top panel is the B.R.$\times dN/dy$ as a function of rapidity for $\mathrm{^3_{\Lambda}H}$ yield in 0-10\% and 10-50\% centrality Au+Au collisions at $\mathrm{\sqrt{s_{NN}}}$ = 3 GeV~\cite{STAR:2021orx}. Bottom panel is the $\mathrm{^3_{\Lambda}H}$ yield at mid-rapidity as a function of $\mathrm{\sqrt{s_{NN}}}$ in central heavy-ion collisions~\cite{STAR:2021orx,ALICE:2015oer}, where symbols and lines represent measurements and theoretical calculations, respectively. Data points at bottom panel assume a B.R. of 25\% for the two-body pionic decay channel from theoretical calculation~\cite{Kamada:1997rv}.}
\end{figure}

Production yields of light nuclei and light hypernuclei are suggested to be related to their internal nuclear structure as well as $N$-$N$ or $Y$-$N$ interaction~\cite{E864:1999zqh,Steinheimer:2012tb,ZhangL}. To examine the structure and the production mechanism, measurements of $\mathrm{^3_{\Lambda}H}$ yield at mid-rapidity ($dN/dy$) have been performed in Au+Au collisions~\cite{STAR:2021orx}, $p$+Pb and Pb+Pb collisions~\cite{ALargeIonColliderExperiment:2021puh,ALICE:2015oer}. Since the decay branching ratio (B.R.) of $\mathrm{^3_{\Lambda}H \rightarrow ^3{He} + \pi^-}$ was not directly measured, experimental data were presented by the $dN/dy$ times the B.R.. The up panel of Fig.~\ref{fig:yield-data} shows the B.R.$\times dN/dy$ of $\mathrm{^3_{\Lambda}H}$ in Au+Au collisions at $\mathrm{\sqrt{s_{NN}}}$ = 3 GeV~\cite{STAR:2021orx}. To interpret the data, calculations from the transport model, JET AA Microscopic Transportation Model (JAM)~\cite{Nara:1999dz} is used to model the dynamical stage of the reaction~\cite{STAR:2021orx}. The coalescence prescription, which is well formulated in studying the cluster formation in heavy-ion collisions~\cite{ExHIC:2011say,Liu:2017rjm,Shao:2020lbq,Hillmann:2021zgj,Zhang:2021ygs,Zhu:2022dlc}, is subsequently applied to the produced hadron as an afterburner~\cite{STAR:2021orx}. In this calculation~\cite{Nara:1999dz,Liu:2019nii}, deuterons are formed through the coalescence of proton and neutron, and subsequently, $\mathrm{^3_{\Lambda}H}$ is formed through the coalescence of $\Lambda$ with deuteron. The coalescence takes place if the relative momentum of the deuteron and $\Lambda$ is within a sphere of radius $p_C$ and the spacial coordinates is within a sphere of radius of $r_C$. It is found that calculations with coalescence parameters ($p_C, r_C$) of (0.12 GeV/c, 4 fm) reproduce the centrality and rapidity dependence of the $\mathrm{^3_{\Lambda}H}$ yield in Au+Au collisions at $\mathrm{\sqrt{s_{NN}}} = $ 3 GeV reasonable well, showing as the dashed curves in Fig.~\ref{fig:yield-data}. This implies that the coalescence mechanism is able to explain the measured data. Together with the $\mathrm{^4_{\Lambda}H}$ yield measurement in Ref.~\cite{STAR:2021orx}, the data offer quantitative input on the coalescence parameters for hypernuclei formation in the high baryon density region, enabling more accurate estimations of the production yields of exotic strange objects~\cite{Steinheimer:2012tb}.

The study of mid-rapidity yields for central collisions over a broad collisions energy will shed more light on the light hypernuclei production mechanism in heavy-ion collisions because the energy density, or more related the effective degree of freedom of the system created in $\mathrm{\sqrt{s_{NN}}}=$ 3 GeV~\cite{STAR:2021yiu} is distinguished different from those created at TeV energy~\cite{Muller:2012zq}. In addition to the aforementioned coalescence model, thermal model is another approach extensively adopted to describe the nucleosynthesis in heavy-ion collisions~\cite{Andronic:2010qu}. Current thermal models do not make any assumption on the structure of the hypernucleus and treat it as a point like particle~\cite{Andronic:2010qu}. However, the dependence of the production on the system configuration is very similar to the coalescence model~\cite{Hillmann:2021zgj}. Bottom panel of Fig.~\ref{fig:yield-data} shows the total $dN/dy$ as a function of center-of-mass energy from a few GeV to a few TeV, with a B.R. of 25\% from theoretical calculation for the $\mathrm{^3_{\Lambda}H \rightarrow ^3{He} + \pi^-}$~\cite{Kamada:1997rv}. By comparing the yields with the theoretical calculations, one sees that canonical ensemble thermal model~\cite{Andronic:2010qu} can approximately describe the $\mathrm{^3_{\Lambda}H}$ data over a few orders of magnitude of $\mathrm{\sqrt{s_{NN}}}$. It is note that canonical ensemble thermal statistics is mandatory to account for the large $\phi/K^-$ and $\phi/\Xi^-$ ratios measured at low energy~\cite{STAR:2021hyx}. Coalescence calculation by the Ultra-Relativistic Quantum Molecular Dynamics Model (UrQMD)~\cite{Bleicher:1999xi} is consistent with the $\mathrm{^3_{\Lambda}H}$ yield at Au+Au collisions of $\mathrm{\sqrt{s_{NN}}}=$ 3 GeV, whereas the UrQMD-hydro hybrid model underestimates the yield at Pb+Pb collisions of $\mathrm{\sqrt{s_{NN}}}=$ 2.76 TeV~\cite{Petersen:2008dd,Zhao:2018lyf}. This may reflect the fact that baryon-antibaryon annihilation is largely different from collisions of a few GeV to a few TeV, due to the longer evolution time in the hadronic stage at higher collision energies. Production yield of $\mathrm{^3_{\Lambda}H}$ also depends on its structure in the coalescence calculation, including the processes $p+n+\Lambda \rightarrow \mathrm{^3_{\Lambda}H}$ and $d+\Lambda \rightarrow \mathrm{^3_{\Lambda}H}$, the yield of $\mathrm{^3_{\Lambda}H}$ is found to be enhanced by about a factor of two compared to the calculation only considering process $p+n+\Lambda \rightarrow \mathrm{^3_{\Lambda}H}$~\cite{Zhang:2018euf}. Calculations from Parton-Hadron-Quantum-Molecular-Dynamics (PHQMD)~\cite{Aichelin:2019tnk,Glassel:2021rod}, which utilizes a dynamical description of hypernuclei formation approach, is consistent with the measured yield within uncertainties. It is noted that measurement of the production yield of $\mathrm{^3_{\Lambda}H}$ in $p$+Pb collision at $\mathrm{\sqrt{s_{NN}}}$ = 5.02 GeV has been reported recently~\cite{ALargeIonColliderExperiment:2021puh}. The measured $dN/dy$ leads to the exclusion with a significance larger than 6$\sigma$ of some configurations of the thermal statistical model~\cite{ALargeIonColliderExperiment:2021puh}. The measured $\mathrm{^3_{\Lambda}H}/\Lambda$ ratio is well described by the two-body coalescence prediction~\cite{Sun:2018mqq} while disfavoured the three-body formulation~\cite{ALargeIonColliderExperiment:2021puh,Sun:2018mqq}.
It is stated in the summary of Ref.~\cite{ALargeIonColliderExperiment:2021puh} that it remains to be seen if advanced version of thermal statistical model using the S-matrix approach account for the interactions among hadrons~\cite{Cleymans:2020fsc} will be able to describe the data.  

Looking at both panels of Fig.~\ref{fig:yield-data}, it is worth mentioning that baryonic interactions in PHQMD are modeled by density dependent two-body baryonic potentials while JAM model adopts a baryonic mean-field approach. Since $\mathrm{^3_{\Lambda}H}$ is a fragile object, it is mostly likely formed at later time of the collisions where the density is low enough that the objects formed are not immediately destroyed. Thus, experimental and theoretical studies on the production yields not only provide quantitative input on the production mechanisms of loosely bound objects in heavy-ion collisions, but may also give information on the time evolution of the high baryon density medium formed~\cite{Braun-Munzinger:2018hat}.

\begin{figure}[htbp]
\includegraphics[width=0.50\textwidth]{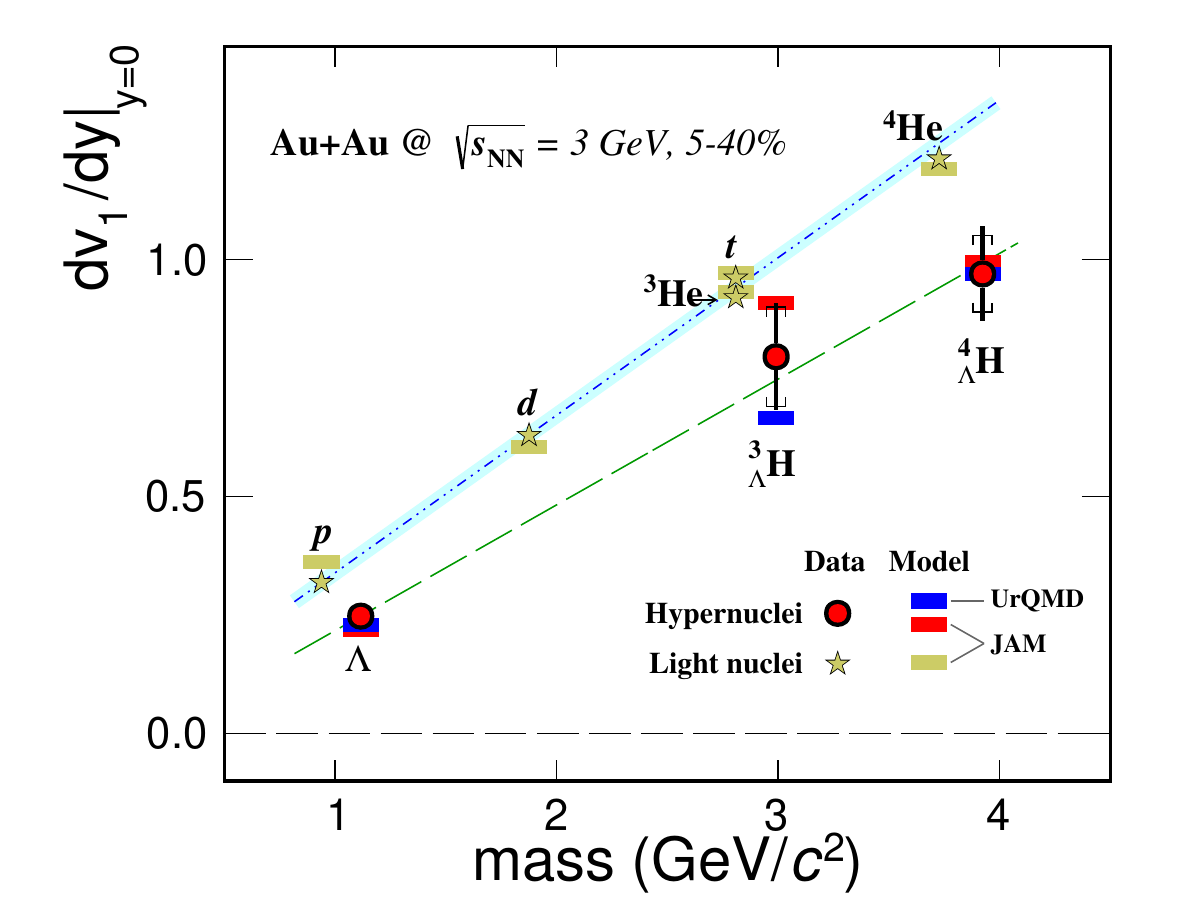}
\caption{\label{fig:v1-data}(Color online) Mass dependence of the mid-rapidity $v_1$ slope, the $dv_1/dy$ for $\Lambda$, $\mathrm{^3_{\Lambda}H}$ and $\mathrm{^4_{\Lambda}H}$ from mid-central Au+Au collisions at $\mathrm{\sqrt{s_{NN}}}=$ 3 GeV~\cite{STAR:2022fnj}. Symbols represent measurements, while boxes represent transport model calculations.}
\end{figure}

Another interesting observable is the excitation function of hypernuclei collective flow, which may provide valuable information for understanding the in-medium $Y$-$N$ interaction~\cite{STAR:2022fnj,Nara:2016phs,Nara:2016hbg,Ma_NST}. It will also provide the possibility to build a connection between heavy-ion collision and the equation-of-state of high baryon density matter which governs the inner structure of compact stars~\cite{Stoecker:2004qu}. Collective flow is driven by the pressure gradient created in such collisions, which has been commonly used for studying the properties of nuclear matter created in collisions~\cite{Ollitrault:1992bk,Ma_FDU,Wang:2022fwq}. Experimentally, it is convenient to quantify flow by the Fourier coefficient of the particle distribution in emission azimuthal angle, measured with respect to the reaction plane. We focus on the first harmonic of the distribution, the so-called directed flow ($v_1$). The word "directed" comes from the fact that such flow looks like a sideward bounce of the fragments away from each in the reaction plane.

Figure~\ref{fig:v1-data} shows the slope of the $dv_1/dy$ near mid-rapidity for $\mathrm{^3_{\Lambda}H}$, together with $\Lambda$ and $\mathrm{^4_{\Lambda}H}$ as a function of particle mass in 5-40\% Au+Au collisions at $\mathrm{\sqrt{s_{NN}}}$ = 3 GeV~\cite{STAR:2022fnj}. Measurements on $p$, $d$, $t$, $^3$He and $^4$He in the similar kinematic window to $\Lambda$ and light hypernuclei have also been performed and plotted for comparison~\cite{STAR:2022fnj}. The slopes $dv_1/dy$ of hypernuclei are systematically lower than those of light nuclei with the same mass number. Linear fits are performed on the mass dependence of $dv_1/dy$ for both light nuclei and hypernuclei. The obtained slope parameters are $0.3323 \pm 0.0003$ for light nuclei and $0.27 \pm 0.04$ for hypernuclei, respectively. Within uncertainties, it seems that the mass dependence of the slope of hypernuclei $v_1$ is similar, but not compatible to that of light nuclei. Calculations from transport models (JAM and UrQMD) plus coalescence as afterburner are in agreement with data within uncertainties~\cite{STAR:2022fnj}. These observations suggest that coalescence of nucleons and $\Lambda$ could be the dominant mechanism for the $\mathrm{^3_{\Lambda}H}$ production in Au+Au collisions at $\mathrm{\sqrt{s_{NN}}}$ = 3 GeV. The measurement may open up a new direction for studying $Y$-$N$ interaction under finite pressure~\cite{Ma_NST}.

Recently, new results on precision measurements of $\Lambda$-$p$ elastic scattering from Jefferson Lab~\cite{CLAS:2021gur} and $\Sigma^-$-$p$ elastic scattering from J-PARC~\cite{J-PARCE40:2021bgw} became available, which may help to constrain the equation-of-state of high density matter inside a neutron star. Systematic studies of momentum correlation functions between proton and hyperons in $pp$ or $AA$ collisions at RHIC or LHC energies provide valuable information to
constrain the $Y$-$N$ interactions~\cite{ALICE:2021njx,STAR:2018uho}. Such as the measurement on the interaction of p$\Lambda$ pairs from zero relative momentum up to the opening of N$\Sigma$ channel yields to a weaker N$\Sigma$$\rightarrow$N$\Lambda$ coupling would require a more repulsive three-body NN$\Lambda$ interaction for a proper description of the hyperon in-medium properties, which has implications on the nuclear equation-of-state and for the presence of hyperons inside neutron stars~\cite{ALICE:2021njx}. Similar study has been preformed in electron-positron collisions. The $\Xi^{0}n \rightarrow \Xi^{-}p$ is observed by using huge $J/\Psi$ events collected with the BESIII detector operating at the BEPCII storage ring~\cite{BESIII:2023clq}. This is the first study of $Y$-$N$ interaction in electron-positron collisions, and opens up a new direction for such research~\cite{Yuan:2021yks}.

\section{Discussions}
There have been several discussions around the $\tau(\mathrm{^{3}_{\Lambda}H})$ measurements since the 1960s~\cite{1964.PR.136.B1803,1966.ILNC.46.786}. The topic is of considerable interest especially in view of the short values, such as the $\tau(\mathrm{^3_{\Lambda}H})$ = ($90_{-40}^{+220}~ps$) from Ref.~\cite{1964.PR.136.B1803} in comparison with the $\tau_{\Lambda}=263.2 \pm2.0~ps$~\cite{ParticleDataGroup:2022pth}. It is learnt that lifetime measurements using visual detectors are notoriously difficult because hypernuclei are in general produced with low kinetic energies and the majority are brought to rest before decay. Determinations of the lifetimes depend on samples in which only very few decay in flight are challenging. Nevertheless, there are nuclear emulsion measurements for the lifetime of light hypernuclei including $\mathrm{^{3}_{\Lambda}H}$ and a helium bubble chamber measurement for $\mathrm{^{3}_{\Lambda}H}$. From the results shown in Fig.~\ref{fig:lifetime-data}, one sees that some early measurements~\cite{1964.PR.136.B1803,1970.NPB.16.46} indicate that $\mathrm{^{3}_{\Lambda}H}$ has a shorter lifetime than the free $\Lambda$.  In particular, the lifetime measurement from Ref.~\cite{1970.NPB.16.46} was $\tau(\mathrm{^3_{\Lambda}H})$ = ($128_{-26}^{+35}~ps$), which presented the smallest error among similar studies in the 1960s and 70s, and was shorter than the others. This measurement was based on the three-body decay channel $^3_\Lambda$H $\rightarrow p + d + \pi^-$ in a nuclear emulsion experiment. The shorter lifetime was attributed to the dissociation of the lightly-bound $\Lambda$ and deuteron when they are traveling in a dense medium~\cite{Bohm:1970av}. However, this explanation is not very satisfying, since measurements performed in Refs.~\cite{1968.PRL.20.819,1969.PR.180.1307,1973.NPB.67.269} also used nuclear emulsion, yet their measurements were close to the $\tau_{\Lambda}$. In addition, Ref.~\cite{1964.PR.136.B1803} used a helium bubble chamber that should not be affected by the hypothesized dissociation, had the lifetime values lower than the free $\Lambda$. Thus, Prof. Dalitz and Prof. Davis commented that the issue was not addressed yet~\cite{Davis:2005mb,Dalitz:2005mc}. Recent heavy-ion experiments have been carried out to address this issue. Measurements of $\tau(\mathrm{^3_{\Lambda}H})$ from experiments at the Relativistic Heavy Ion Collider (STAR Collaboration), at the SIS18 (HypHI Collaboration), and at the Larger Hadron Collider (ALICE Collaboration) were reported~\cite{STAR:2010gyg,Rappold:2013fic,ALICE:2015oer,STAR:2017gxa,ALICE:2019vlx,STAR:2021orx,ALICE:2022rib}. These data showed the $\tau(\mathrm{^3_{\Lambda}H})$ value $\approx$ (0.5 - 0.96) $\tau_{\Lambda}$, presenting a similar large spread of the $\tau(\mathrm{^3_{\Lambda}H})$ values, as observed in the older nuclear emulsion and helium bubble chamber measurements (c.f. Fig.~\ref{fig:lifetime-data}).

On the other hand, there has been much interest in applying effective field theory (EFT) methods to nuclear systems with two or more nucleons. EFTs provide a powerful framework to explore a separation of scales in physical systems in order to perform systematic, model-independent calculations~\cite{Weinberg:1990rz}. Calculation of important quantities in nuclear physics using lattice QCD is becoming practical, usually at the physical values of strange quark but not at the physical value of light-quark masses~\cite{NPLQCD:2012mex}. Closely related to the topic of current paper, calculations~\cite{Kamada:1997rv} based on a $\mathrm{^3_{\Lambda}H}$ wave function and $3N$ scattering states from rigorous solution of three-body Faddeev equations using realistic $N$-$N$ and $Y$-$N$ interactions yielded a lifetime only 3\% smaller than the $\tau_{\Lambda}$. It is worth mentioning that the $\tau_{\Lambda}$ from Ref.~\cite{Kamada:1997rv} is 272 $ps$, which is 3\% larger than the world data~\cite{ParticleDataGroup:2022pth}. A  simple model calculations~\cite{1992.JPG.18.339} by taken the $\Lambda$ to orbit an unperturbed deuteron in a $\Lambda$-deuteron potential based on a separable $\Lambda$-nucleon potential predicted a lifetime $\approx$13\% smaller than the $\tau_{\Lambda}$. In such study, the author found that as long as the binding energy of $\mathrm{^3_{\Lambda}H}$ is reproduced, the lifetime calculation is rather insensitive to the fine details of the particular $\Lambda$$N$ interaction model chosen~\cite{1992.JPG.18.339}. Within a closure-approximation calculation in which the associated exchange matrix element is evaluated with wave functions obtained by solving the $\mathrm{^3_{\Lambda}H}$ three-body Faddeev equation, Ref.~\cite{Gal:2018bvq} had the result $\tau(\mathrm{^3_{\Lambda}H})\approx 0.90 \tau_{\Lambda}$. Introducing $\pi$ final state interaction in terms of $\pi$ distorted scattering waves in Ref.~\cite{Gal:2018bvq} results in $\tau(\mathrm{^3_{\Lambda}H})\approx 0.81 \tau_{\Lambda}$. Different calculations seem to cover the measurements worldwide. A much more precise measurement will be important to pin down the potential different physics involved in different calculations.

\begin{figure}[htbp]
\includegraphics[width=0.50\textwidth]{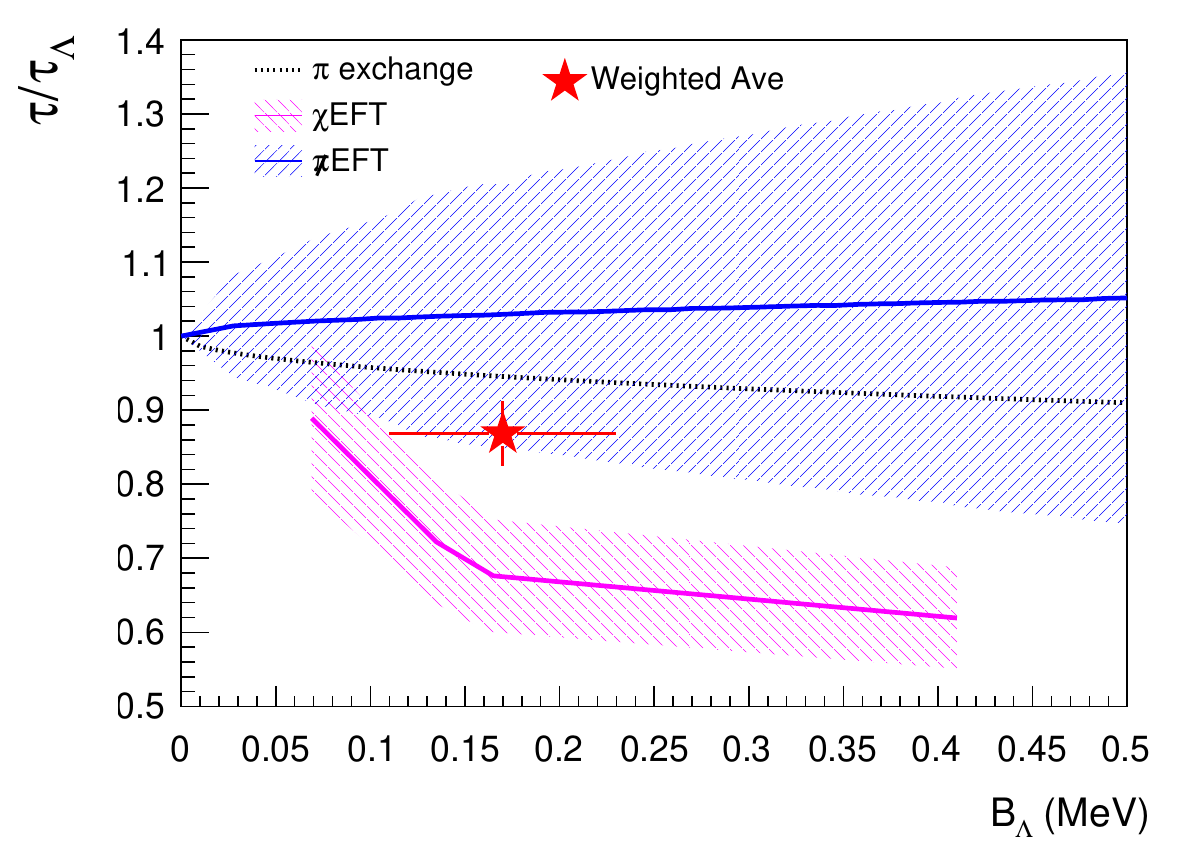}
\caption{\label{fig:lifetime-binding}(Color online) Correlation of the $\mathrm{^3_{\Lambda}H}$ lifetime relative to the free $\Lambda$ lifetime as a function of the $\Lambda$ separation of $\mathrm{^3_{\Lambda}H}$. Data point is the average of world data, and curves are calculations from $\pi$ exchange~\cite{1966.ILNC.46.786}, pionless effective field theory~\cite{Hildenbrand:2020kzu} and chiral effective field theory~\cite{Perez-Obiol:2020qjy}.}
\end{figure}

Since the $\mathrm{^3_{\Lambda}H}$ lifetime and $\Lambda$ separation energy are strongly correlated, we can explore the underlying physics by the $\Lambda$ separation energy measurements. $\pi$ exchange studies using a closure approximation to sum over the final nuclear states reached in the $\mathrm{^3_{\Lambda}H}$ weak decay, reduced the $\mathrm{^3_{\Lambda}H}$ lifetime calculation to the evaluation of a $\mathrm{^3_{\Lambda}H}$ matrix element~\cite{1966.ILNC.46.786,Dalitz:1962eb}. With a choice of the closure energy $q = 96~\rm MeV/c$, Rayet and Dalitz deduced the correlation between lifetime and binding as $\Gamma$($\mathrm{^3_{\Lambda}H}, J=\frac{1}{2}$) = (1+0.14$\sqrt{B_{\rm \Lambda}}$)$\Gamma_{\rm \Lambda}$~\cite{1966.ILNC.46.786}. Calculation based on pionless effective field theory ($\slashed{\pi}$EFT) with $\Lambda$ and deuteron degrees of freedom shows that the sensitivity of the total width to binding energy is small, while the partial widths for decays into individual final states and the ratio $R = \Gamma_{\rm ^3He}/(\Gamma_{\rm ^3He}+\Gamma_{pd})$ is large~\cite{Hildenbrand:2019sgp,Hildenbrand:2020kzu}. That is, the $R$ value will increase quite significantly as $B_{\Lambda}$ increased. It seems that the $R$ quantity appears to be better suited to determine $B_{\Lambda}$ indirectly than the total width~\cite{1973.NPB.67.269,1992.JPG.18.339}. Employing $\mathrm{^3_{\Lambda}H}$ and $\rm ^3He$ three-body wave functions generated by ab initio hypernuclear no-core shell model, microscopic calculation~\cite{Perez-Obiol:2020qjy} of the $\mathrm{^3_{\Lambda}H}$ mesonic decay rate $\Gamma$($\mathrm{^3_{\Lambda}H \rightarrow  ^3He + \pi^-}$) found that : (i) replacing pionic plane wave by realistic $\pi^- - \rm^3He$ distorted wave enhances the decay rate by $\approx$15\%. (ii) The $\Sigma NN$ admixtures in $\mathrm{^3_{\Lambda}H}$ reduce the purely $\Lambda NN$ decay rate by $\approx$10\% due to interference effects. Their calculation also suggests that the $\tau(\mathrm{^3_{\Lambda}H})$ varies strongly with the rather poorly known $\Lambda$ separation energy. One learns that such poorly known $B_{\Lambda}$ serves as a free-parameter entering into the EFT and can be varied without changing other observables~\cite{Hildenbrand:2019sgp,Hildenbrand:2020kzu}. The $B_{\Lambda}$ is one of the key quantities to discriminate different theoretical scenarios~\cite{Le:2019gjp}.

Figure~\ref{fig:lifetime-binding} shows the correlation of $\tau(\mathrm{^3_{\Lambda}H})$ and $B_{\Lambda}$ of the $\mathrm{^3_{\Lambda}H}$. The pink curve represents the $\chi$EFT calculation with different harmonic oscillator basis ultraviolet scales~\cite{Perez-Obiol:2020qjy}. The color black curve is from $\pi$ exchange calculation~\cite{1966.ILNC.46.786} while the color blue is the result from $\slashed{\pi}$EFT~\cite{Hildenbrand:2020kzu}. Bands are uncertainties of each calculations. The $\chi$EFT is able to associate different measurements of $\tau(\mathrm{^3_{\Lambda}H})$ with its own underlying values of $B_{\Lambda}$~\cite{Perez-Obiol:2020qjy,Gazda:2023fow}. We here plot the world average values from Fig.~\ref{fig:lifetime-data} and Fig.~\ref{fig:bl-data} instead of putting each individual measurement. We see on Fig.~\ref{fig:lifetime-binding} that the average value is consistent with calculations from $\pi$ exchange theory, $\slashed{\pi}$EFT, $\chi$EFT considering the data uncertainties and the calculation uncertainties.

There had been discussions on the partial mesonic decay rate to the total rate mainly because the branching ratio for various decay modes of a hypernucleus will generally depend on both the spin of the hypernucleus and the nature of the $\Lambda$ decay interaction~\cite{Leon:1959zz,Dalitz:1959zz}. From experiment side, it is challenging to measure the absolute decay rate but more straight forward to measure the ratio of different decay modes by conducting different decay modes in the same experiment. For the $\mathrm{^3_{\Lambda}H}$, this ratio is defined as $R_3 = \frac{\Gamma(\mathrm{^3_{\Lambda}H \rightarrow ^3He + \pi^-})}{\Gamma(\mathrm{^3_{\Lambda}H \rightarrow ~all~\pi^-~channels})}$. It had been measured by different experiment from early days~\cite{1968.PRL.20.819,1973.NPB.67.269,Bertrand:1970nm} and the modern heavy-ion collisions~\cite{STAR:2017gxa}, with all data being consistent with the average value of $R_3 = 0.35 \pm 0.04$, and being consistent under the assumption $J(\mathrm{^3_{\Lambda}H})= \frac{1}{2}$ within data uncertainties. On-going measurement with uncertainty less than 10\% is progressing well and may provide stringent constrains in model calculation~\cite{Leung:2023gki}.

\section{Summary}
We discuss measurements of lifetime, $\Lambda$ separation energy, production yields and collective flow on the lightest hypernuclei, the $\mathrm{^3_{\Lambda}H}$ to gain knowledge of the $Y$-$N$ interaction.
Both lifetime and $\Lambda$ separation energy data show a large spread among different measurements, thus we perform a statistical analysis to include all data. The average values are $\tau(\mathrm{^3_{\Lambda}H}) = 228.5 \pm 11.6~ps$, $B_{\Lambda}= 0.17 \pm 0.06~\rm MeV$, while the $\tau(\mathrm{^3_{\Lambda}H}) = 236.4 \pm 8.1~ps$ is an average analysis of measurements in heavy-ion collisions. Those values serve as input parameters to constrain the effective field theory calculations. Considering the large uncertainties in the measurements and theoretical calculations, different scenarios of calculations can describe the average values. Data on production yields and directed flow are well described by the transport plus coalescence afterburner calculations, in which interactions among nucleons and strange baryons are important ingredients. Data on $\mathrm{^3_{\bar{\Lambda}}\bar{H}}$ to $\mathrm{^3_{\Lambda}H}$ mass difference is consistent with zero within $10^{-4}$ precision. Independent and on-going measurements with better precision will be helpful to nail down the possible systematic and shed light on the structure of the $\mathrm{^3_{\Lambda}H}$ and the $Y$-$N$ interaction. Further theoretical development for hypernucleus is also expected.

\section{Perspective}

In the last two decades, the relativistic heavy-ion collisions have proved to be an effective way of producing the simplest hypernucleus and studying its properties since such measurement~\cite{E864:2002xhb} at Alternating Gradient Synchrotron (AGS) was first performed in the early 2000s several decades after the last experimental emulsion measurements of $\mathrm{^3_{\Lambda}H}$ in the 1970s. 
In the last decade, the improved detector capabilities and increased statistics at RHIC and the LHC facilities have offered unique opportunities for precise lifetime and $\Lambda$ separation energy measurements. The precision of lifetime measurement has significantly improved to be $\pm10$ ps while the $\Lambda$ separation energy measurement has been hovering around $50-100$ keV with an improved control and understanding of the systematical uncertainties. To date, the world-average lifetime measurement shows a 3-$\sigma$ significance lower than the free $\Lambda$ lifetime with a precision of $\pm10$ ps. With the improved detector capability and increased high-statistical data of ALICE Detector at the LHC~\cite{ALICE:2022rib} and STAR Detector at RHIC~\cite{STAR:2021orx}, this would be further improved likely by another factor of 2 in the near future. A1 at MAMI~\cite{A1:2022erj} using novel high-luminosity lithium target and E07 at J-PARC~\cite{Nakagawa:2022xyz} using machine learning on analyzing emulsion data are promising in producing the most precise measurements of the $\Lambda$ separation energy with a precision of $\pm20-30$ keV in the near future. New measurement of hypernucleus lifetime using in-flight $\rm ^4He$$(K^-,\pi^0)$$\mathrm{^4_{\Lambda}H}$ reaction has obtained $\tau(\mathrm{^4_{\Lambda}H}) = 206 \pm 8(\rm stat.) \pm 12(\rm syst.)~ps$, one of the most precise measurements to date~\cite{Akaishi:2023nqx}. They are preparing to measure the lifetime of $\mathrm{^3_{\Lambda}H}$ using the same setup in the near future~\cite{Akaishi:2022esy}.
These should provide measurements of a non-zero value with more than 5-$\sigma$ significance if the nominal values of the world-average $\Lambda$ separation energy and lifetime (different from $\Lambda$) hold. The endeavor in the recent decades would finally provide solid experimental input for further theory developments. 


\section{Acknowledgements.}
We thank S. Wang on the help for designing Fig.~\ref{fig:h3l-structure}. The work of J. Chen was supported in part by the National Key Research and Development Program of China under Contract No. 2022YFA1604900, by the National Natural Science Foundation of China (NSFC) under Contract No. 12025501. The work of Y.-G. Ma was supposed in part by the NSFC under Contract No. 11890710, 11890714, and 12147101. The work of X. Dong and Z. Xu was funded by the U.S. DOE Office of Science under contract No. DE-sc0012704, DE-FG02-10ER41666, DE-AC02-98CH10886, and DE-KB0201022.

\bibliographystyle{unsrt}
\bibliography{ref-h3l}

\end{document}